\documentclass[letter]{myaa}
\usepackage{txfonts}
\usepackage{natbib}
\usepackage[a4paper,breaklinks]{hyperref}
\bibpunct{(}{)}{;}{a}{}{,}
\usepackage{graphicx}
\usepackage{balance}
\usepackage{latexsym}
\def\kms {\,$\mathrm{km\, s^{-1}}$}
\def\ms {\,$\mathrm{m\, s^{-1}}$}
\newcommand{\cobold}{\ensuremath{\mathrm{CO}^5\mathrm{BOLD}}}
\newcommand{\linfor}{Linfor3D}
\newcommand{\Teff}{\ensuremath{T_\mathrm{eff}}}

\idline{1}{1}
\begin{document}
\title{Line shift, line asymmetry,  and the $^6$Li/$^7$Li isotopic ratio determination 
\thanks{Based on 
observations carried out at the European
Southern Observatory (ESO), under prog. ID 75.D-0600.
}}
\author{ Roger Cayrel\inst{1}
\and Matthias Steffen \inst{2}
\and Hum Chand\inst{3}
\and  Piercarlo Bonifacio\inst{4,5,6}
\and Monique Spite\inst{4}
\and Fran\c{c}ois Spite\inst{4}
\and Patrick Petitjean\inst{3}
\and Hans-G{\"u}nter Ludwig \inst{4,5}
\and Elisabetta Caffau \inst{4}
} 
\authorrunning{Cayrel et al.}
\titlerunning{Line shift, line asymmetry and $^6$Li/$^7$Li}
\offprints{R. Cayrel} 
\institute{GEPI, Observatoire de Paris, CNRS, Universit\'e Paris Diderot; 61 av. de l'Observatoire, F-75014 
Paris,  France
\and  Astrophysikalisches Institut Potsdam, An der Sternwarte 16, D-14482
Potsdam, Germany
\and Institut d'Astrophysique de Paris, 98 bis bd Arago, F-75014 Paris, France 
\and
GEPI, Observatoire de Paris, CNRS, Universit\'e Paris Diderot; Place
Jules Janssen 92190
Meudon, France
\and  CIFIST, Marie Curie Excellence Team
\and
Istituto Nazionale di Astrofisica,
Osservatorio Astronomico di Trieste,  Via Tiepolo 11,
I-34143 Trieste, Italy
}
\date{Received / Accepted}

\abstract
{Line asymmetries are generated by   
convective Doppler shifts
in stellar
atmospheres, especially in metal-poor stars, where convective motions
penetrate to higher atmospheric levels. Such asymmetries are usually neglected
in abundance analyses.
The determination of the $^6$Li/$^7$Li isotopic ratio
is prone to suffering from such asymmetries, as the contribution of $^6$Li is 
a slight blending 
reinforcement of the red wing of each component of the corresponding
$^7$Li line, with respect to its blue wing.} 
{The present paper studies the halo star HD 74000
and estimates the impact of convection-related
asymmetries on the 
Li isotopic ratio determination. 
}
{Two methods are used to meet this aim. The first, which is 
purely empirical, consists  in deriving
a template profile from another element that can be assumed to originate in
the same stellar atmospheric layers as \ion{Li}{i},
producing absorption lines of approximately 
the same equivalent width as individual 
components of the \ion{$^7$Li}{i}  resonance line.
The second method consists in conducting the abundance analysis based on
NLTE line formation 
in a 3D hydrodynamical model atmosphere, taking into account the effects of 
photospheric convection.}
{The results of the first method show that the convective asymmetry
generates an excess absorption in the red wing of the $^7$Li absorption feature that mimics
the presence of $^6$Li at a level comparable to the hitherto 
published values.
This opens the possibility that only an upper limit on $^6$Li/$^7$Li 
has thus far been derived.
The  second method confirms these findings.}
{From this work, it appears that a systematic reappraisal of
former determinations of $^6$Li abundances in halo stars is warranted.}
\keywords{Hydrodynamics - Line: profiles - Stars:abundances - Stars: Population II - Stars:individual: HD 74000 }      
\maketitle
\section{Introduction}
Hydrodynamical simulations 
of stellar atmospheres,  including an {\em ab initio} treatment 
of convection,
have reached a stage where line profiles are accurately predicted from the 
Doppler shifts induced by convective motions
(Stein and Nordlund  \citealt{SN98}; Asplund et al. \citealt{ANTAPS00}). 
The present state of  the art 
is described in the third  
edition of the book ``The Observation and Analysis of 
Stellar Photospheres'' (chapter 17) by David Gray \citeyearpar{gray05}.
The effect of convection, 
through Doppler shift, on the shape of the atmospheric 
absorption lines,  is to create asymmetric profiles.
This has no consequence on abundance determination from isolated lines
since a numerical integration 
over the profile gives the correct equivalent width. 
Problems arise
when one is interested in an unresolved blend. 
This is the case for the measurement of $^6$Li/$^7$Li 
isotopic ratios, which makes use of
the $^6$Li + $^7$Li  blend, where 
both the presence of $^6$Li and the line asymmetry caused by
convection result in the reinforcement of the red wing of the $^7$Li feature.
Curiously, the consequences of the line asymmetry induced in 
turn-off (TO) metal-poor halo stars, in which most detections of 
$^6$Li have been made, have not been studied in detail. 
Since $^6$Li, unlike $^7$Li, is not (or barely) formed in
the standard big bang nucleosynthesis, its presence
in metal-poor stars requires a suitable production channel.
From a different perspective, the presence of $^6$Li in metal-poor
stars poses strong constraints on any theory of Li depletion
in such stars, since $^6$Li is destroyed by nuclear reactions
at lower temperatures than $^7$Li.
Proposals of $^6$Li production mechanisms to explain the
observations in Pop II stars include: decaying massive particles
at the time of nucleosynthesis, variations of the fundamental
physical constants, structure formation shocks and solar-like flares.
The implications of some of these
are far reaching and it is therefore  
of great importance to make the measurements
as robust as possible. 
The study of the impact of line asymmetries
on measured Li isotopic ratios is here
performed, first observationally, from an empirical point 
of view, and second, theoretically, by computing the $^6$Li/$^7$Li blend from 
a 3D hydrodynamical model atmosphere in order to corroborate the empirical
result. We have  chosen to perform our observational investigation on
the TO metal-poor star HD~74000 
because 
only two upper limits to 
its  $^6$Li/$^7$Li ratio exist
\citep{ht97,smith98} 
and 
because this star lies 
extremely close, in the HR-diagram, to
HD 84937, the archetype of a star with
measurable $^6$Li.  Twenty one-hour 
exposures of HD~74000 were 
obtained with HARPS \citep{harps} in service mode at 
ESO La Silla (Chile).
We combined individual exposures into a single 
spectrum after realignment in wavelength compensating for the Earth's motion. 
The resulting spectrum has a  S/N ratio of 600 per pixel 
and a spectral resolution of 120,000. 
Section 2 deals with the observational study, 
whereas Sect. 3 is a 3D-NLTE study of the 
formation of the 6708~\AA\ $^6$Li+$^7$Li blend, essentially to 
investigate the 
asymmetry in the line profile induced by the hydrodynamic motions. 
This aids in
understanding the empirical results. 
Section 4 gives the conclusions of the work. 
\section{Line asymmetry derived from observation}
 \begin{figure}
\centering
  \resizebox{0.8\hsize}{!}{\includegraphics[clip=true]{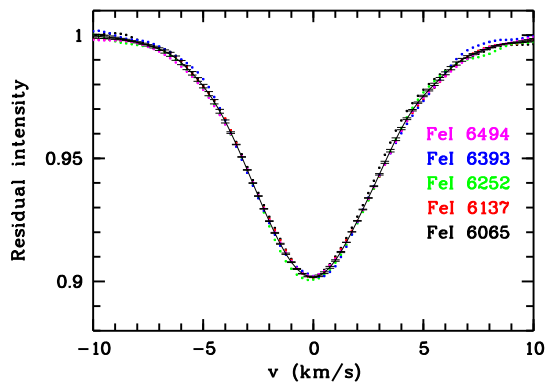}}
\caption{The profiles of the five selected iron lines after rescaling. Their
mean (solid line) and the median profiles are practically identical.}
 \label{fivelines}
 \end{figure}
\begin{figure}
\centering
 \resizebox{\hsize}{!}{\includegraphics[clip=true]{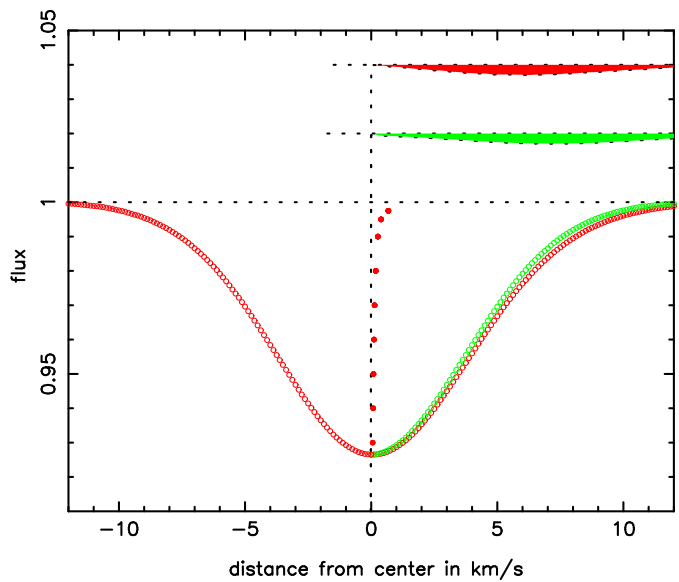}}
 \caption{Asymmetry expected on the 
main component of the $^7$Li doublet from the 
mean profile of the 5 iron lines of Fig.~\ref{fivelines}. 
The asymmetric profile is represented 
by red circles. The symmetric profile 
of the blue wing is represented 
by green circles.
The excess absorption in the true profile
compared to the symmetric red wing is shown with the
continuum shifted to 1.04 for clarity.
The effect of a $^6$Li blend corresponding to a
$^6$Li/$^7$Li of 4\%  
is also shown, with  the continuum at 1.02, as reference.   
The two features are strikingly similar, both  in size and position.
The full red dots mark the line bisector.
 \label{asymmetryobs}
}
\end{figure}
\begin{figure}
\centering
\resizebox{0.8\hsize}{!}{\includegraphics[clip=true]{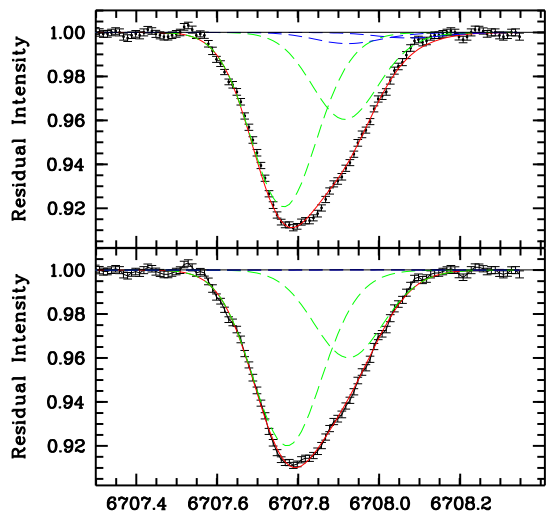}}
\caption{Upper panel: 
The observed Li doublet (black dots with error bars), together with the best 
fitting synthetic profile (red solid line)
computed using the empirical profile
derived from the \ion{Fe}{i} lines. The green dashed lines
show the $^7$Li components and the blue dashed lines the $^6$Li
components. Fitting parameters are the contributions
 of $^7$Li and $^6$Li and thermal broadening, but  no global wavelength shift
was allowed. See text for details.
Lower panel:
like upper panel, but also allowing
for a global wavelength shift, the best fit corresponds
to a shift of 372 \ms.
}
 \label{partialfit}
 \end{figure}
\begin{figure}[h!]
\centering
\resizebox{0.8\hsize}{!}{\includegraphics[clip=true]{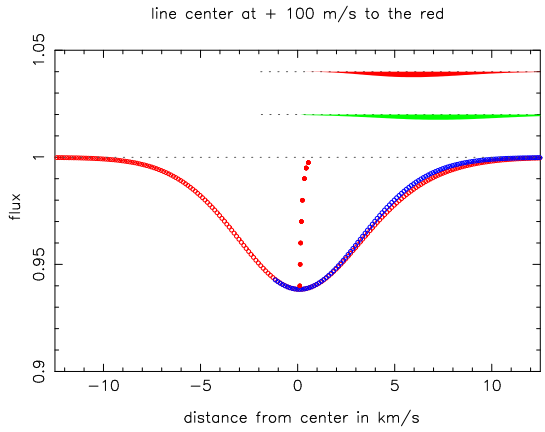}}
\caption{Profile of a single component of $^7$Li computed  with Linfor3D in
  LTE. It is based on the average over 11 snapshots from a  \cobold\ model
  atmosphere run with T$_\mathrm{eff}$  =6375 K , log(g)=4.0, 
[Fe/H]=-2.0. The asymmetry is characterised by the difference
between the red wing 
(circles) and the mirror image of the blue wing (blue circles)
The area between both is shown in red at the top, as in Fig. 2.
The asymmetry is 3 \% in area, and the convective 
line shift is 100 m/s to the red.}   
\label{3DLTE}
\end{figure}

\begin{figure}[h!]
\centering
\resizebox{0.8\hsize}{!}{\includegraphics[clip=true]{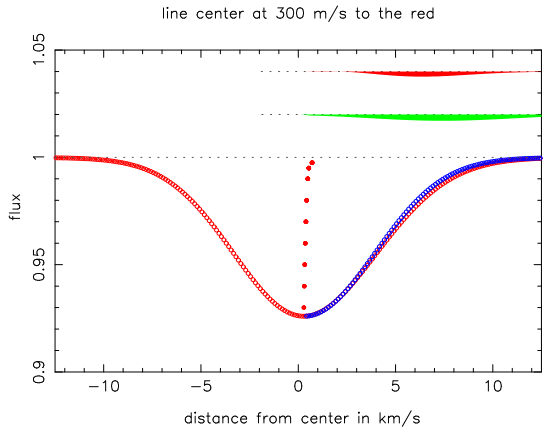}}
\caption{Like Fig.~\ref{3DLTE} but for the case of 3D-NLTE. The 
convective shift is 
3 times larger and the area asymmetry slightly less.}
\label{3DNLTE}
\end{figure} 

The goal is to 
derive a template profile from single component absorption lines 
that are assumed to arise in the same atmospheric layers 
as the \ion{Li}{i} lines and
have similar characteristics. 
In particular, the lower level of the transition should be at about 
5.392~eV from the ionisation level. 
Several \ion{Fe}{i}  lines satisfy this constraint and have, in addition, 
similar line equivalent widths as the main $^7$Li component of the blend, 
i.e.  
about 16~m\AA\ accounting for 2/3 of
the total equivalent width of the Li feature,
measured
to be  26~m\AA\ in HD 74000. 
We have 
added a third constraint on the 
wavelength; itt should not be too far from that of the
Li blend so as
to have a similar continuous opacity. Wavelengths shorter than
6700~\AA\ were favoured since the fringing is increasing towards the
red.  Table~1 lists the five 
chosen transitions, their wavelength, excitation energy,
observed full width half-maximum (FWHM) and equivalent width.
\par
 We started by convolving the five
 individual profiles by a Gaussian profile with FWHM~=~1~\kms\ (or about
 22~m\AA) slightly larger than the pixel size (17~m\AA). 
This reduces the photon noise without significant loss of
 resolution (the isotopic separation of $^6$Li and $^7$Li is 0.16~\AA\ or
 7.15~\kms). The absorption lines are then rebinned on a velocity scale with
zero velocity fixed at the place of maximum line 
depth and a step of 0.25~\kms. 
The line centre is obtained by a Gaussian fit to the line core.
The depressions are  finally scaled linearly  to the same 
central value. 
Figure~\ref{fivelines} gives the result of the scaling for the five
selected iron lines.
 The variation of the continuous opacity of H$^-$ between 6000~\AA\ and
 6700~\AA\ is very small (about 1\%) and does not significantly affect the
Fe~{\sc i} and Li~{\sc i} absorptions.
To combine the profiles, we allow for variations in central wavelengths and 
line depths.
The final template 
profile was determined by a simultaneous least square fit of the 
scaling and shifts of the five lines. 

\begin{figure}
\begin{center}
\resizebox{0.8\hsize}{!}{\includegraphics[clip=true]{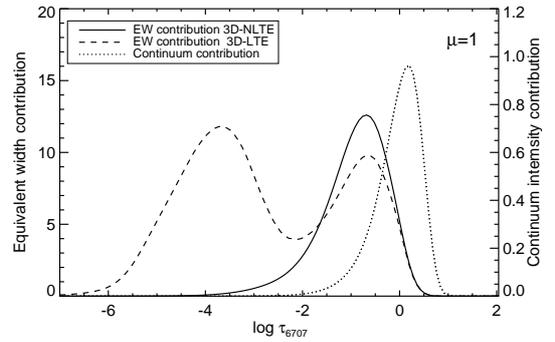}}
\end{center}
\vspace{-0.5cm}
\caption{The  contribution functions of the (disk-centre) 
equivalent width of the \ion{Li}{i}
6707.8\AA\ line (solid lines, left ordinate) and of the (disk-centre) 
continuum intensity (dotted line,
right ordinate) in 3D-LTE  and 3D-NLTE. 
Note the double peak in the equivalent width contribution
function in 3D-LTE for the Li line. 
The peak at the surface is due to an increased LTE population of the levels 
at the low  surface temperatures reached in the 3D model.
In 3D-NLTE, the departures from LTE completely annihilate 
this peak and  strengthen
the inner peak.}      
\label{contrib}
\end{figure}

Before discussing more specifically the $^6$Li/$^7$Li ratio,  we have
to adapt the profile derived from Fe~{\sc i} to Li~{\sc i}. Indeed,
due to different atomic weights, the Li~{\sc i} lines are broader.
For a temperature of 5500~K, the line-of-sight thermal broadening
$\sqrt{(kT/m)}$ is 0.9~\kms\ for $^{56}$Fe,  2.55~\kms\ 
for $^7$Li, and 2.76~\kms\ for
$^6$Li.
The profile derived from Fe~{\sc i} lines must  be convolved with a Gaussian
profile of 
width 2.39~(resp. 2.61) \kms\ (5.6, resp. 6.1 \kms\ FWHM) to be adapted to the  
$^7$Li (resp. $^6$Li) components of the Li blend. Figure \ref{asymmetryobs}
shows this convolved profile (red circles).

The asymmetry of the mean profile is usually described by the so-called
bisector, which is the locus of the mid-points of
horizontal line segments 
joining the two points of the profile 
at equal residual intensity, on the blue and red wings. 
For a symmetric line, the bisector is a straight vertical line.  
 
In Fig.~\ref{asymmetryobs} we illustrate the asymmetry of our derived 
mean profile by plotting the mirror image of the left wing with respect to 
the vertical line at the centre. 
It is apparent that an excess absorption exists in the red wing.
The difference between the red wing and the mirror of the blue wing is shown 
with the continuum shifted to 1.04. For comparison, we show 
the depression of a line representing 4\% of the
main line equivalent width, with the continuum shifted to 1.02 (see figure).
The quasi-degeneracy between these 
two signals is the central problem we address 
in this paper.
  If one ignores the asymmetry in the modelling of the absorption feature
(which is the case in all 1D-LTE analyses), 
one will overestimate the strength of the component blended in the
red wing, and hence derive a spurious $^6$Li abundance.

To compute an empirical profile of the \ion{Li}{i} resonance
doublet we assumed that the 
individual line components (doublet, isotopic, hyperfine)
have the same shape as the observed mean iron profile. We further assumed
that they are located at the same wavelengths and have the same $gf$-values
as used by \citet{asplund}. 
Each component was weighted by its $gf$-value and 
the abundance of $^6$Li or $^7$Li. 
Our fitting parameters were therefore the two abundances,
an additional thermal broadening and a global
velocity shift of the whole absorption profile
to allow for uncertainties in the wavelength calibration.

Figure \ref{partialfit} 
displays the results of our fits to
the observed resonance \ion{Li}{i} doublet
in HD 74000 (see also Table~\ref{tabfit}). 
We note that scanning of the $\chi^2$ in parameter space, as 
well as Monte Carlo simulations, 
show that isotopic ratio, velocity shift and thermal broadening
are correlated parameters. 
The fit in which all the parameters are left free requires
an emission of the $^6$Li component.
This is, of course, 
unphysical and must be interpreted as evidence that when the
empirical asymmetric profile is 
used there is no need for a contribution by $^6$Li.
In the upper panel,  the velocity shift is fixed to zero,
the low  goodness of fit
in this case makes the fit, at most, marginally
acceptable.
On the other hand, it can be seen from Table ~\ref{tabfit}
that a fit with $^6$Li fixed to zero is good, 
confirming that the assumption of the absence of
$^6$Li allows an acceptable fit.
To assess the impact of the line asymmetry on the derived
$^6$Li/$^7$Li ratio, 
we constructed a symmetric profile
by mirroring 
the blue wing of our empirical profile, 
which matches the unblended part of the $^7$Li component.
This results in  a good fit (not displayed here), but requires 
an isotopic ratio of 2.3\%.

Our  conclusion, based on the use of the
asymmetric empirical profile,  is that 
there is no positive detection of  $^6$Li  in HD~74000.

\section{Effects of line asymmetries from theoretical 3D-NLTE simulations}
\begin{figure}
\centering
\resizebox{0.8\hsize}{!}{\includegraphics[clip=true]{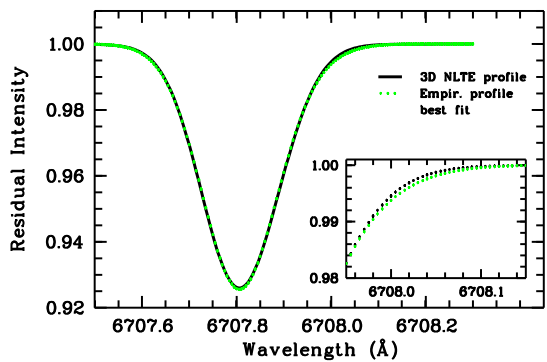}}
\caption{Comparison between empirical and 3D-NLTE profile.}
\label{comp}
\end{figure}%
\begin{figure}
\centering
\resizebox{0.8\hsize}{!}{\includegraphics[clip=true]{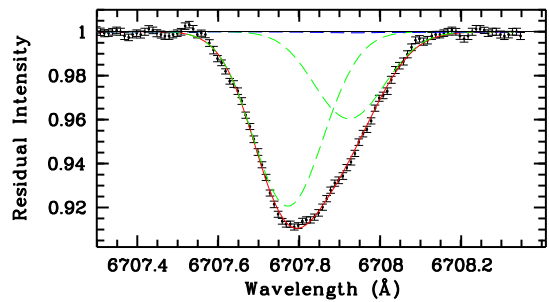}}
\caption{Fit with 3D-NLTE profile providing $^6$Li/$^7$Li = 0.6\%.}
\label{fitrot}
\end{figure}

To corroborate our empirical findings, 
we performed a theoretical investigation
of the line formation process based on a 3D hydrodynamical model atmosphere.
Spectrum synthesis calculations of the Li feature were 
first performed
assuming LTE, and subsequently by
applying a 3D-NLTE code to study the impact of
departures from LTE on the resulting line profile.

A synthetic spectrum was 
calculated with the code  \linfor\ developed in Potsdam and 
Meudon\footnote{\href{http://www.aip.de/~mst/Linfor3D/linfor_3D_manual.pdf}{http://www.aip.de/$\sim$mst/Linfor3D/linfor\_3D\_manual.pdf}}.
The spectrum is based  on an average over 11 independent snapshots 
(each representing 28$\times$28 Mm$^2$ of the stellar surface with
140$\times$140 grid points)
generated by the radiation-hydrodynamics code \cobold\ 
\citep{freytag02,wedemeyer04}. 
The atmospheric parameters of the model are \Teff = 6375
log g = 4.0 and metallicity --2.0, as shown 
in Table~\ref{tabatm}, together with a selection of parameters
found in the literature
The
result is shown in Fig.~\ref{3DLTE}. The asymmetry is very similar in shape
and size to the one obtained from the empirical observational study. 

We have repeated the same computation using the same snapshots, applying this
time a 3D-NLTE code newly developed by two of us (M. Steffen and R. Cayrel). 
 The code uses an 8 level \ion{Li}{i} model atom including 11 transitions 
and the corresponding radiative processes, electron 
collisions and the Li+H charge
exchange process computed by Barklem et al. \citeyearpar{BBA03}.
Figure \ref{3DNLTE} shows the resulting line profile, slightly less asymmetric
than the one obtained  under the 3D-LTE assumption, but still exhibiting an 
asymmetry equivalent to a
$^6$Li/$^7$Li ratio blend of 2.7\%, comparable to  the value derived in our empirical study.

A completely new aspect has been brought to light by the 3D-NLTE study: there is
a drastic change of the contribution function to the equivalent
width in 3D-NLTE with respect to 3D-LTE 
(see Fig.~\ref{contrib}). The contribution of the
very cool surface layers has been removed
by departure coefficients of about 1/100, 
reducing the equivalent width of
the full $^6$Li+$^7$Li feature from 59~m\AA\ to 23~m\AA\ for the standard
abundance of $^7$Li of 2.2 (log(nH)=12 scale). Linked to this change, the
$^7$Li and $^6$Li line central wavelengths have been redshifted by 7 m\AA, or
about 300 \ms.  This explains why, during a quarter of a century, 
it has always been
necessary to shift the synthetic feature to the red of its rest-wavelength,
a shift largely above the uncertainty in wavelength calibrations
with high resolution spectrographs.

Figure~\ref{comp} illustrates the close correspondence between our empirical
(iron line based) Li line profile and the synthetic 3D-NLTE profile. In
Fig.~\ref{fitrot} we show the result of our best fit using the 3D-NLTE profile
as a building-block of the Li feature. 
The theoretical profile has the correct thermal broadening. We have 
rotationally broadened the theoretical profile by 1~\kms, as described in
\citet{ludwig}, to match the projected rotational velocity of the star. 
Note that  the required additional
broadening is 2.66~\kms, which is very close to the nominal resolution of
HARPS. We 
obtain a $^6$Li/$^7$Li ratio of 0.6\% and 
estimate the error on the Li isotopic ratio to be 2\%.  A Monte
Carlo simulation, taking into account only Poisson noise, provides an error
estimate of 1.6\%.  Our estimated total uncertainty of 2\% takes into account
the presence of non-Poisson noise (e.g. residual fringing).

We have computed the asymmetry for LTE and NLTE with a 
gravity $\log(g)$ =4.5 (see Table~\ref{tabatm}), instead of 4.0, 
and found very different results, 
with an asymmetry 3 times smaller,
showing that the asymmetry is very gravity dependent. 
Obviously computations covering a 
substantial fraction of the parameters,$\mathrm{T_{eff}}$, 
g and metallicity, are needed.
\balance

\section{Conclusions} 

We provide evidence
for a significant degree of degeneracy  between the asymmetry
caused by hydrodynamical effects on the \ion{Li}{i} resonance line, and the
asymmetry caused by the presence of a $^6$Li blend. From both an observational
study of the spectrum of HD 74000, and 3D-LTE and 3D-NLTE line
synthesis computations, we conclude that the size and shape of the
convection-related asymmetry is equivalent to a contribution of $^6$Li by a
few percent.   We find in HD74000, 
$^6$Li/$^7$Li $=(0\pm2)$\%.  Moreover,
the 3D-NLTE approach allows us to understand the 6 to 10~m\AA\ shift
commonly needed to fit the observed profile with a 1D synthetic profile.

 Strictly taken, our results do not imply that previous measurements of
the $^6$Li/$^7$Li isotopic ratio are invalidated since  $^6$Li has not
been previously
detected in HD74000. However, we systematically find smaller
isotopic ratios when fitting the observed spectrum with asymmetric 
line profiles 
rather than 
symmetric ones. This 
is what
we expect from the influence of the line asymmetry, but
opposite to what \citet{asplund} obtain.
  The reason for this discrepancy is unclear, but we believe that our study
  features an unprecedented accuracy by treating line asymmetries in a largely
  model-independent fashion, working with a spectrum of outstanding
  S/N ratio, spectral resolution and wavelength accuracy, as well
  as including NLTE effects in the 3D modelling.  Therefore, a serious
  reappraisal of the $^6$Li abundances derived from earlier works seems
  warranted.

\Online

\begin{table*}[h!]
\begin{minipage}[t]{\columnwidth}
\caption{ List of selected \ion{Fe}{i} lines.}
\centering
\renewcommand{\footnoterule}{}  
\begin{tabular}{cccccl}
\hline\hline
 Mult.\footnote{Nave et al. 1994} & wavelength & $\chi_\mathrm{exc}$ & 
$\chi_\mathrm{ion} - \chi_\mathrm{exc}$ &
FWHM & EW \\
&  (\AA)      & (eV)     & (eV)     &(\kms)& (m\AA)\\
\hline
536 & 6065.4822 & 2.608 & 5.294 & 0.129 & 13.1 \\
536 & 6137.6920 & 2.588 & 5.314 & 0.136 & 17.5 \\
465 & 6252.5554 & 2.404 & 5.498 & 0.135 & 14.7 \\
464 & 6393.6013 & 2.433 & 5,469 & 0.140 & 17.0 \\
464 & 6494.9805 & 2.404 & 5.498 & 0.142 & 30.4\footnote{Although stronger than the other 
lines this scales linearly and provides a higher S/N ratio.} \\
\hline
\end{tabular}
\end{minipage}
\end{table*}

\begin{table*}[h!]
\begin{minipage}[t]{\columnwidth}
\caption{Results of the fit to the observed \ion{Li}{i} doublet of HD 74000.\label{tabfit} }
\renewcommand{\footnoterule}{}  
\centering
\begin{tabular}{lrrrr}
\hline\hline
Type of fit & $^6$Li/$^7$Li\footnote{a negative value indicates that the 
fitted $^6$Li line is emission.}
 & Shift\footnote{This is the fitted shift necessary to match the observed profile,
note that while the empirical and 1D profiles have a zero intrinsic shift the 3D profiles have
+100\ms and +300\ms convective shifts respectively. } & Broadening & Goodness of fit\footnote{Defined as the probability 
of obtaining a $\chi^2$ as high as the one observed.}  \\
            &     \%          & \ms  & \kms       &                  \\
\hline
Empirical Prof. & --2.8\phantom{$^d$}    & 371\phantom{$^d$}     & 5.81       & 0.304\\
Empirical Prof. &   6.3\phantom{$^d$}    &  0\footnote{fixed}   & 5.15       & 0.002\\
Empirical Prof. &   0$^d$          & 372\phantom{$^d$}   & 5.77       & 0.316\\
Symmetr. Prof.  &  2.3\phantom{$^d$}     & 412\phantom{$^d$}     & 5.92       & 0.272\\
1D LTE          & 2.4\phantom{$^d$}      & 394\phantom{$^d$}     & 5.80       & 0.233 \\
3D LTE\footnote{rotationally broadened by 1 \kms}
           & 0.2\phantom{$^d$}    & 290\phantom{$^d$}     & 4.29       & 0.324 \\
3D NLTE$^e$          &  0.6\phantom{$^d$}    &  82\phantom{$^d$}     & 2.66      & 0.303 \\
\hline

\end{tabular}
\end{minipage}
\end{table*}

\begin{table*}[h!]
\caption{Atmospheric parameters of  HD 74000.\label{tabatm} }
\begin{tabular}{rrrr}
\hline\hline
\Teff & log g & [Fe/H] & Ref\\
 6375 &4.00&-2.00 & \cobold\ model\\
 6320 &4.50&-2.00 & \cobold\ model\\
 6126 &4.01&-1.80 & Zhang \& Zhao 2005\\
 6040 &4.20&-2.17 & Arnone et al. 2005\\
 6392 &4.27&-1.96 & Melendez \& Ramirez 2004\\
 6109 &    &-1.96 & Nordstr\"om et al. 2004\\
 6203 &4.03&-2.05 & Gehren et al. 2004\\
 6216 &4.09&-1.96 & Gratton et al. 2003\\
 6025 &4.10&-2.0  & Fulbright 2000\\
 6040 &    &-2.02 & Ryan et al. 1999 \\
 6190 &4.13&-2.00 & Smith et al. 1998\\
 6090 &4.00&-2.00 & Hobbs \& Thorburn 1997\\
 6184 &4.13&-1.69 & Nissen et al. 1997\\
 6224 &4.50&-2.00 & Alonso et al. 1996\\
 6341 &5.19&-1.52 &Gratton et al.  1996\\
 6090 &4.15&-2.08 &Beveridge \& Sneden 1994\\
 6211 &3.92&-1.93 &Axer et al. 1994\\
 6090 &4.35&-1.98 &Tomkin et al. 1992\\
 6146 &3.90&-2.23 &Magain 1989\\
 6072 &4.2 &-2.06 &Hartmann \& Gehren 1988\\
 6222 &4.5 &-1.80 &Peterson 1978\\

\hline

\end{tabular}
\end{table*}


\begin{thebibliography}{}

\bibitem[(2000)]{ANTAPS00}
Asplund, M., Nordlund, \AA, Trampedach, R. et al. 2000, A\& A 888,555
\bibitem[Asplund et al.(2006)]{asplund} Asplund, M., Lambert, 
D.~L., Nissen, P.~E., Primas, F., \& Smith, V.~V.\ 2006, \apj, 644, 229

\bibitem[(2003)]{BBA03}
Barklem, P.S., Belyaev, A.K.,\& Asplund, M., 2003, A \& A, 409, L1

\bibitem[Freytag et al.(2002)]{freytag02} Freytag, B., Steffen, 
M., \& Dorch, B.\ 2002, AN, 323, 213

\bibitem[Gray(2005)]{gray05} Gray, D.~F.\ 2005, The 
Observation and Analysis of Stellar Photospheres, 3rd Edition, by 
D.F.~Gray.~ ISBN 
0521851866.
Cambridge, UK: Cambridge University Press, 2005.

\bibitem[Hobbs \& Thorburn(1997)]{ht97} Hobbs, L.~M., \& 
Thorburn, J.~A.\ 1997, \apj, 491, 772 

\bibitem[Ludwig (2007)]{ludwig}
Ludwig, H.-G.,\ 2007, \aap, in press, arxiv:0707.3347

\bibitem[Mayor et al.(2003)]{harps} Mayor, M., et al.\ 2003, 
The Messenger, 114, 20 


\bibitem[Nave et al.(1994)]{nave} Nave, G., 
et al.\ 1994, \apjs, 94, 221


\bibitem[Smith et al.(1998)]{smith98} Smith, V.~V., Lambert, 
D.~L., \& Nissen, P.~E.\ 1998, \apj, 506, 405 

\bibitem[(1998)]{SN98}
Stein, R.F. \& Nordlund, \AA, 1998 ApJ 499, 914  


\bibitem[Wedemeyer et al.(2004)]{wedemeyer04} Wedemeyer, S., 
et al.\ 2004, \aap, 414, 
1121 



\end{thebibliography}

\clearpage

\begin{thebibliography}{}

\bibitem[Alonso et al.(1996)]{1996A&AS..117..227A} Alonso, A., Arribas, S., 
\& Martinez-Roger, C.\ 1996, \aaps, 117, 227 

\bibitem[Arnone et al.(2005)]{2005A&A...430..507A} Arnone, E., Ryan, S.~G., 
Argast, D., Norris, J.~E., \& Beers, T.~C.\ 2005, \aap, 430, 507 

\bibitem[Axer et al.(1994)]{1994A&A...291..895A} Axer, M., Fuhrmann, K., \& 
Gehren, T.\ 1994, \aap, 291, 895 

\bibitem[Beveridge \& Sneden(1994)]{1994AJ....108..285B} Beveridge, R.~C., 
\& Sneden, C.\ 1994, \aj, 108, 285 

\bibitem[Fulbright(2000)]{2000AJ....120.1841F} Fulbright, J.~P.\ 2000, \aj, 
120, 1841 

\bibitem[Gehren et al.(2004)]{2004A&A...413.1045G} Gehren, T., Liang, 
Y.~C., Shi, J.~R., Zhang, H.~W., \& Zhao, G.\ 2004, \aap, 413, 1045 

\bibitem[Gratton et al.(1996)]{1996A&A...314..191G} Gratton, R.~G., 
Carretta, E., \& Castelli, F.\ 1996, \aap, 314, 191 

\bibitem[Gratton et al.(2003)]{2003A&A...404..187G} Gratton, R.~G., 
Carretta, E., Claudi, R., Lucatello, S., \& Barbieri, M.\ 2003, \aap, 404, 
187 

\bibitem[Hobbs \& Thorburn(1997)]{ht97} Hobbs, L.~M., \& 
Thorburn, J.~A.\ 1997, \apj, 491, 772 

\bibitem[Hartmann \& Gehren(1988)]{1988A&A...199..269H} Hartmann, K., \& 
Gehren, T.\ 1988, \aap, 199, 269 

\bibitem[Magain(1989)]{1989A&A...209..211M} Magain, P.\ 1989, \aap, 209, 
211 


\bibitem[Mel{\'e}ndez \& Ram{\'{\i}}rez(2004)]{2004ApJ...615L..33M} 
Mel{\'e}ndez, J., \& Ram{\'{\i}}rez, I.\ 2004, \apjl, 615, L33 

\bibitem[Nissen et al.(1997)]{1997ESASP.402..225N} Nissen, P.~E., Hoeg, E., 
\& Schuster, W.~J.\ 1997, Hipparcos - Venice '97, 402, 225 

\bibitem[Nordstr{\"o}m et al.(2004)]{2004A&A...418..989N} Nordstr{\"o}m, 
B., et al.\ 2004, \aap, 418, 989 

\bibitem[Peterson(1978)]{1978ApJ...222..181P} Peterson, R.\ 1978, \apj, 
222, 181 

\bibitem[Ryan et al.(1999)]{1999ApJ...523..654R} Ryan, S.~G., Norris, 
J.~E., \& Beers, T.~C.\ 1999, \apj, 523, 654 

\bibitem[Smith et al.(1998)]{smith98} Smith, V.~V., Lambert, 
D.~L., \& Nissen, P.~E.\ 1998, \apj, 506, 405 

\bibitem[Tomkin et al.(1992)]{1992AJ....104.1568T} Tomkin, J., Lemke, M., 
Lambert, D.~L., \& Sneden, C.\ 1992, \aj, 104, 1568 



\bibitem[Zhang \& Zhao(2005)]{2005MNRAS.364..712Z} Zhang, H.~W., \& Zhao, 
G.\ 2005, \mnras, 364, 712 
\end{thebibliography}
\end{document}